\documentclass[10pt,preprint]{emulateapj} 
\def\msun{\rm M_{\sun}} 

\def\lsun{\rm L_{\sun}} 
\def\micron{$\mu$m} 

\def\mdot{\rm \dot{M}}

\begin{document}

\shortauthors{Espaillat et al.} 
\shorttitle{CVSO 224}

\title{A slowly accreting $\sim$10 Myr old transitional disk in Orion
OB1a}

\author{C. Espaillat\altaffilmark{1}, J. Muzerolle\altaffilmark{2}, J.
Hern\'{a}ndez\altaffilmark{1,3}, C. Brice{\~n}o\altaffilmark{3},  N.
Calvet\altaffilmark{1},   P. D'Alessio\altaffilmark{4}, M.
McClure\altaffilmark{5}, D. M. Watson\altaffilmark{5}, L. Hartmann\altaffilmark{1}, 
\& B. Sargent\altaffilmark{5}}

\altaffiltext{1}{Department of Astronomy, University of Michigan, 830
Dennison Building, 500 Church Street, Ann Arbor, MI 48109, USA;
ccespa@umich.edu, ncalvet@umich.edu, hernandj@umich.edu,
lhartm@umich.edu} 
\altaffiltext{2}{Steward Observatory, University of
Arizona, Tucson, AZ 85712, USA; jamesm@as.arizona.edu}
\altaffiltext{3}{Centro de Investigaciones de Astronom\'{i}a(CIDA),
Merida, 5101-A, Venezuela; briceno@cida.ve} 
\altaffiltext{4}{Centro de
Radioastronom\'{i}a y Astrof\'{i}sica, Universidad Nacional Aut\'{o}noma
de M\'{e}xico, 58089 Morelia, Michoac\'{a}n, M\'{e}xico;
p.dalessio@astrosmo.unam.mx} 
\altaffiltext{5}{Department of Physics and
Astronomy, University of Rochester, NY 14627-0171, USA;
melisma@astro.pas.rochester.edu, dmw@astro.pas.rochester.edu}

\begin{abstract}

Here we present the {\it Spitzer} IRS spectrum of CVSO 224, the sole
transitional disk located within the $\sim$10 Myr old 25 Orionis
group in Orion OB1a.  A model fit to the spectral energy distribution of
this object indicates a $\sim$7 AU inner disk hole that contains a small
amount of optically thin dust.  In previous studies, CVSO 224 had been
classified as a weak-line T Tauri star based on its H${\alpha}$ equivalent
width, but here we find an accretion rate of 7$\times$10$^{-11}$
${\msun}$ yr$^{-1}$ based on high-resolution Hectochelle data.
CVSO 224's low $\mdot$ is in line with photoevaporative clearing
theories. However, the {\it Spitzer} IRS spectrum of CVSO 224 has
a substantial mid-infrared excess beyond 20{\micron} which indicates
that it is surrounded by a massive outer disk.  Millimeter measurements
are necessary to constrain the mass of the outer disk around CVSO 224 in
order to confirm that photoevaporation is not the mechanism behind
creating its inner disk hole.

\end{abstract}


\section{Introduction}

Stars surrounded by transitional disks have characteristics that fall
between objects that have clear evidence for disks and stars with no
disk material.  They have deficits of infrared flux at $\lambda$ $<$ 8 {\micron} but
show strong excesses at longer wavelengths, indicating that the
innermost regions have undergone significant clearing of small dust grains
\citep{strom89, skrutskie90}.

The {\it Spitzer Space Telescope} has greatly improved our resolution in
the infrared and has recently provoked extensive modeling studies of
several transitional disks around T Tauri stars e.g TW Hya
\citep{uchida04}, GM Aur, DM Tau
\citep{calvet05}, CS Cha \citep{espaillat07a}, HD 98800B
\citep{furlan07}, and Hen 3-600 \citep{uchida04}.  These objects have
been explained as inwardly truncated disks which are optically thick to
the stellar radiation, with most of the mid-infrared emission
originating in the inner edge or ``wall" of the outer disk.  Spectral
energy distributions (SEDs) point to holes in these disks of less than
$\sim$40 AU.  The inner holes of the transitional disks around DM Tau and
CoKu Tau$/$4 are cleared of small dust grains \citep{dalessio05, calvet05}
while the transitional disks GM Aur, TW Hya, CS Cha, HD 98800B, 
and Hen 3-600 have
a small yet detectable near-infrared excess produced by some
submicron and micron sized optically thin dust remaining in the inner
disk hole \citep{calvet05, calvet02, espaillat07a,furlan07,uchida04}. In addition
to these disks with holes (i.e. large reductions of small dust grains from the
star out to an outer optically thick wall), the ``pre-transitional disk"
class has recently been identified \citep{espaillat07b}.  These objects,
exemplified by UX Tau A and LkCa 15, have an inner optically thick disk
separated from an outer optically thick disk by an optically thin gap
\citep{espaillat07b, espaillat08}.

One possible cause for these dust clearings in transitional and
pre-transitional disks is planet formation.
Hydrodynamical simulations have shown that a newly formed planet could
accrete and sweep out the material around it through tidal disturbances
\citep{quillen04, rice03, paardekooper04, varniere06}.  Stellar
companions could also dynamically clear the inner disk
\citep{artymowicz94}.  Additional dust clearing theories include
inside-out evacuation induced by the MRI \citep{chiang07} and
photoevaporation \citep{clarke01}.

The holes of GM Aur, TW
Hya, and DM Tau can potentially be explained by the MRI.
It has been proposed that the MRI can operate on the ionized inner wall
of the disk which leads material to accrete from the wall onto the star,
creating a hole in the disk that grows from the inside-out
\citep{chiang07}.  
In order for photoevaporation to be effective, the disk mass must be below 
$\sim$0.005 M$_{\sun}$ and the disk accretion rate must fall below 
the wind rate ($\sim$10$^{-10}$ M$_{\sun}$ yr$^{-1}$; Alexander \& Armitage
 2007; AA07). 
Photoevaporation cannot explain the holes of these
three objects since their outer disks are massive (M$_{disk}$ $>$  0.05 M$_{\sun}$; Calvet et al. 2002, 2005) 
and their accretion rates are higher than the
photoevaporative wind rate.  
Neither the MRI nor photoevaporation
can explain the gaps seen in UX Tau A and LkCa 15 given that these
inside-out clearing mechanisms cannot account for a remnant optically
thick inner disk. Photoevaporation could in principle explain CoKu Tau$/$4's hole
since this object has a weak outer disk (M$_{disk}$ $\sim$  0.0005 M$_{\sun}$; Andrews \& Williams 2005) and a non-detectable accretion rate, however, its disk clearing is
most likely caused by its recently discovered stellar companion
\citep{ireland08}. HD 98800B and Hen 3-600 are additional examples of a 
transitional disk where a stellar companion is apparently responsible for the
disk's hole \citep{furlan07,uchida04}.  \citet{guenther07} also found a companion
in the inner hole of CS Cha although the separation of the stellar pair
is less than 5 AU (Eike Guenther, private communication) which is too
small to explain the truncation of the outer disk at $\sim$ 43 AU
\citep{espaillat07a} with existing models \citep{artymowicz94}.  To
date, a radial velocity detection of a planetary companion in a transitional disk has
only been claimed in TW Hya \citep{setiawan08}.

While most transitional disk studies have focused on objects located in
1--2 Myr old star-forming regions with relatively isolated star
formation, here we present detailed modeling of CVSO 224, the only
transitional disk located in the $\sim$10 Myr old 25 Orionis group
\citep{briceno07} in the Orion OB1a subassociation \citep{briceno05}. 

\section{Observations \& Data Reduction} \label{sec:obs} 
We present the
SED of CVSO 224, which is also known as 1a$\_$1200 (Hern\'{a}ndez et al. 2007) and
05254675+0143303 (2MASS), in Figure 1.  B- and R-band photometry were
taken from the USNO database. V- and I-band photometry \citep{briceno07}
as well as J-, H-, and K-band (2MASS) are shown.  IRAC and MIPS data were
taken from \citet{hernandez07}.

An A$_{V}$ of 0.21 for CVSO 224 is derived from fitting an M3
photosphere (KH95) to the observations, and the data are dereddened
with the \citet{mathis90} reddening law.  The derived A$_{V}$ is
consistent with a mean extinction of 0.5 mag found toward Orion OB1a by
\citet{calvet05orion}.  Stellar parameters (M$_{*}$, R$_{*}$; Table 1)
are derived from the HR diagram and the Baraffe evolutionary tracks
\citep{baraffe02} using a T$_*$ of 3470 K (KH95) for an M3 star
\citep{briceno07}. The distance to 25 Orionis is 330pc
\citep{briceno07}.

CVSO 224 was observed by the $Spitzer$ IRS instrument on March 8, 2006
(AOR ID: 16264960) with the short-wavelength, low-resolution (SL) module
of IRS and the long-wavelength, low-resolution (LL) module, covering
$\sim$5 to 40 {\micron}, at a resolving power of
$\lambda$/$\delta$$\lambda$ = 60-100.  The observation was carried out
in IRS Staring Mode.  We used the Spectral Modeling, Analysis, and
Reduction Tool (SMART) software package developed by the IRS instrument
team \citep{higdon04} to extract and calibrate the spectrum.  \citet{furlan06}
can be consulted for further data-reduction details.

In Figure 2 we show the high-resolution (R$\sim$34,000) spectrum of
CVSO 224 centered on the H$\alpha$ 6563{\AA} line.  It was obtained with the
Hectochelle multifiber instrument (Szentgyorgyi et al. 1998) mounted on
the 6.5m MMT at Mt. Hopkins, Arizona for the dataset presented in
\citet{briceno07}. We refer the reader to that article for details on
the observations and data reduction.

\begin{deluxetable}{l c | l c} 
\tabletypesize{\scriptsize} \tablewidth{0pt}
\tablecaption{Stellar and Model Properties\label{tab:prop}} 
\startdata
\hline 
\hline 
\multicolumn{2}{c}{Stellar Properties} & \multicolumn{2}{c}{Optically Thick Outer Wall} \\ 

\hline 
M$_{*}$(M$_{\sun}$) & 0.3 & a$_{min}$ ({\micron}) & 0.005 \\
R$_{*}$ (R$_{\sun}$) & 0.9 & a$_{max}$({\micron}) & 0.25 \\  
T$_{*}$ (K) & 3470 & T$_{wall}$ (K) & 120\\
L$_{*}$ ($\lsun$) & 0.1 & z$_{wall}$ (AU) & 1\\ 
Distance (pc) & 330 &  R$_{wall}$ (AU) & 7 \\ 
$\mdot$ (M$_{\sun}$yr$^{-1}$) & 7$\times$10$^{-11}$ & &\\ 
Inclination (deg) & 30 & &\\ 
A$_{V}$ & 0.21 & &\\ 
Spectral Type & M3 & &\\ 
\hline 
\multicolumn{4}{c}{Optically Thin Inner Region}\\ 
\hline
a$_{min, thin}$ ({\micron}) & 0.005 & R$_{in, thin}$ (AU) & 0.04 \\ 
a$_{max, thin}$ ({\micron}) & 4 & R$_{out, thin}$ (AU) & 1\\ 
M$_{dust,thin}$ (M$_{\sun}$) & 4$\times$10$^{-12}$  & &
\enddata
\end{deluxetable}

\begin{figure} 
\figurenum{1} 
\label{sedfig} 
\epsscale{0.7}
\plotone{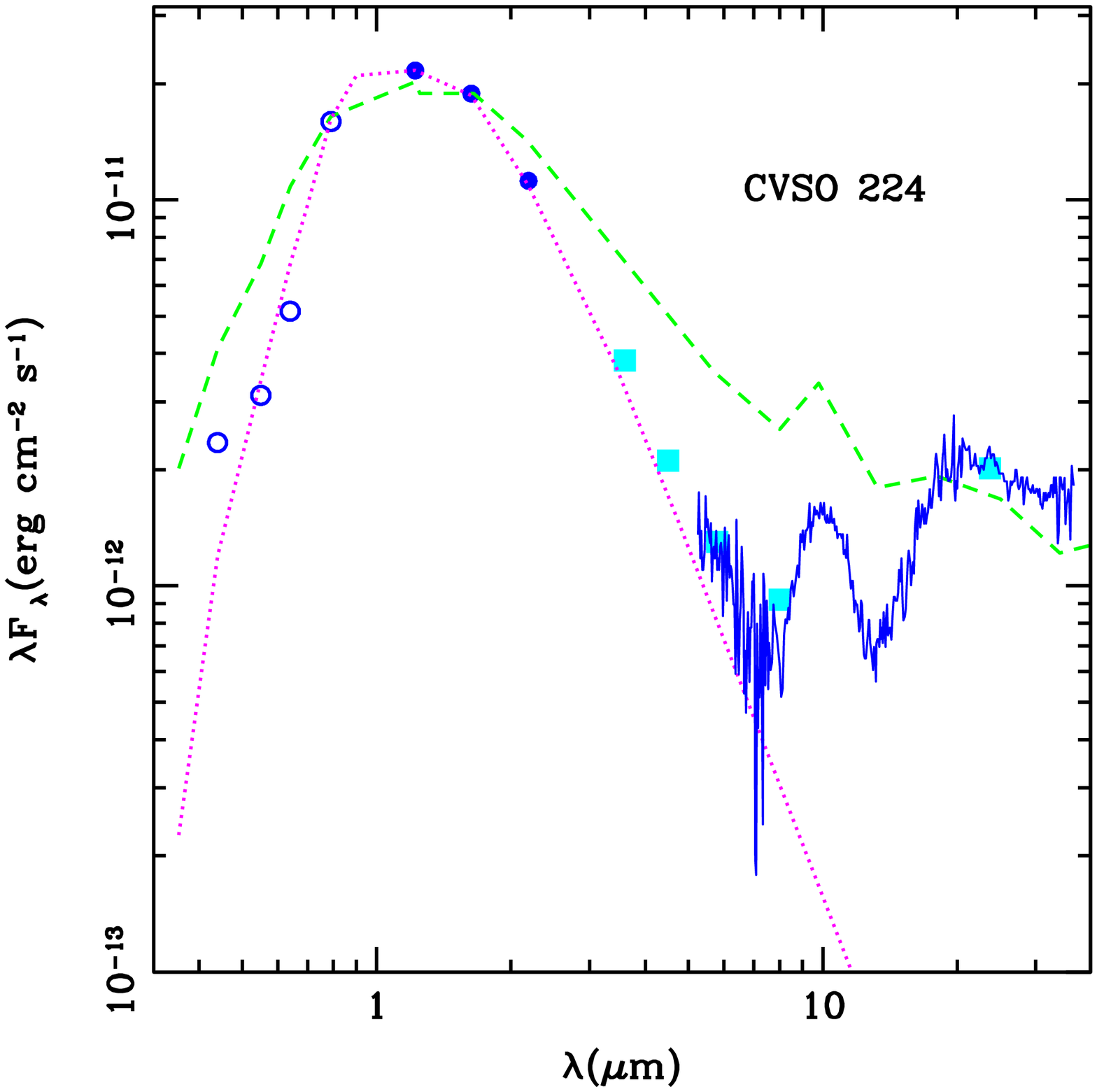} 
\caption{SED of CVSO 224.  We show dereddened
ground-based optical (open circles), J,H,K (filled circles), $Spitzer$
IRAC and MIPS (closed squares), and $Spitzer$ IRS (solid line) data.  We
also show the median SED of Taurus \citep[short-dashed line]{dalessio99,
furlan06}, which represents a full disk, and an M3 photosphere
\citep[dotted line]{KH95}.  [See the electronic edition of the Journal
for a color version of this figure.]
}
\label{figsed} \end{figure}

\begin{figure} 
\figurenum{2} 
\label{mdotfig} 
\epsscale{0.7}
\plotone{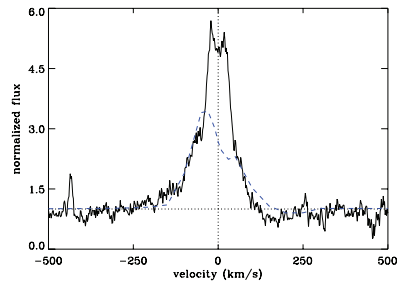} 
\caption{Hectochelle spectrum of CVSO 224 (solid
line) with the best-fit accretion model (dashed line).  We measure an
accretion rate of 7$\times$10$^{-11}$ M$_{\sun}$ yr$^{-1}$.}
\label{figsed} 
\end{figure}

\begin{figure} 
\figurenum{3} 
\label{figmodel} 
\epsscale{0.7}
\plotone{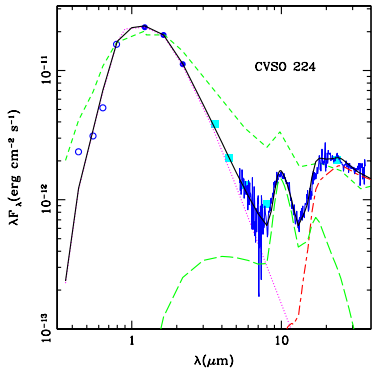} 
\caption{SED and transitional disk model of CVSO
224.  Symbols are the same as in Figure 1.  The best-fit model (heavy solid line)
corresponds to an inner disk hole of 7 AU that has
$\sim$4$\times$10$^{-12}$ M$_{\sun}$ of optically thin dust located
between the dust destruction radius and 1 AU; the disk is relatively
devoid of small dust grains between 1 and 7 AU.  Model components are as
follows: stellar photosphere (dotted line), 
optically thin dust region (long-dash), outer wall
(dot-short-dash).  [See the electronic edition of the Journal for a
color version of this figure.]
}
\label{figsed} \end{figure}

\section{Analysis}

\subsection{Accretion Properties}

According to theories of magnetospheric accretion,
the inner disk is truncated by the stellar magnetic field and material
is channeled onto the star via accretion columns.  This leads to strong,
broad H${\alpha}$ emission profiles due to the high temperatures and
velocities of the accreting material \citep{muzerolle01}. 
\citet{whitebasri03} showed that a star is accreting if EW(H${\alpha}$)
$\geq$ 20 {\AA} for M3 stars.  According to this criteria, CVSO 224, a
M3 star  \citep{briceno07} with EW(H${\alpha}$) of $\leq$ 20 {\AA}, can
be classified as a weakly accreting T Tauri star (WTTS)
\citep{briceno07}.  However, some stars which 
lie just below this EW(H${\alpha}$) limit could be
slow accretors \citep{barrado03}. 
When analyzing the high-resolution spectrum of CVSO 224 taken with
Hectochelle (Fig. 2) we find that the width of the H${\alpha}$ profile
at 10$\%$ is larger than 200 km/s indicating that this star is accreting
\citep{whitebasri03}.  

We estimated the accretion rate using the
magnetospheric accretion models of Muzerolle et al. (2001).  See that
paper for the details of the fitting procedure.  In short, we calculated
models using the mass, radius, and effective temperature of the star as
fixed inputs.  The gas temperature, density (calculated from $\mdot$),
and inclination were then varied to find the best fit to the observed
line profile.  The profile of CVSO 224 has a central self-reversal as
seen in WTTS (e.g. Figure 1 of Muzerolle et al.\ 2000), but also has broad
wings and redshifted absorption.  This suggests that we are seeing signatures
of both the stellar chromosphere and the accreting
material traveling in the magnetospheric field lines.  Since the line
core is dominated by chromospheric emission, which is not included in
the models, only the line wings were considered in the fit.  
The resultant best-fit model parameters are
$i$=30$^{\circ}$, T$_{max}$=12,000 K, $\mdot$=7$\times$10$^{-11}$
M$_{\sun}$ yr$^{-1}$. The inclination angle derived here is constrained
to about 50$\%$ of the nominal value of 30$^{\circ}$ since the actual
geometry is likely much more complicated \citep{muzerolle01}.  We adopt
this value for CVSO 224 since there is no other estimate for the
inclination angle.

\subsection{Disk Properties}

When we compare CVSO 224 to the median SED of Taurus (Fig. 1; D'Alessio et al. 1999, Furlan 
et al. 2006) which has been shown to be representative of a
full disk \citep{dalessio99, dalessio06}, it is apparent that
there is a strong infrared deficit, indicating that CVSO 224 is
surrounded by a transitional disk. We follow D'Alessio et al. (2005) to
calculate the structure and emission of the optically thick disk's inner
edge or ``wall," assumed to be vertical and axisymmetric. The radiative transfer in the wall atmosphere is
calculated with M$_*$, R$_*$, T$_*$, distance,
inclination, and $\mdot$ as well as minimum and maximum grain sizes
(Table 1). We use a grain-size distribution that follows a power-law of
$a$$^{-3.5}$, where $a$ is the grain radius.  We assume ISM sized grains
and adopt a$_{min}$=0.005 {\micron} and a$_{max}$=0.25 {\micron}
\citep{drainelee84}. The wall has a radial gradient of temperature and
we use its outermost temperature, T$_{wall}$, as a free parameter in
fitting the SED.  This temperature, combined with the dust composition,
determines the wall radius \citep{dalessio05}.  The best fit to the SED
is shown in Figure 3 and corresponds to T$_{wall}$=120 K and
R$_{wall}$=7 AU.  The wall height, z$_{wall}$ (1 AU), is also a
free parameter which is dependent on the best-fit to the SED.  Varying
the inclination angle within 50$\%$ of 30$^{\circ}$ 
does not change the size of the inner hole.

CVSO 224 has a near-infrared excess; the ratio of its observed to photospheric
emission 
at 5.8 {\micron} is 1.6.
The low value of T$_{wall}$ implies that the wall can account for
neither the 10 {\micron} silicate-feature emission nor the small near-IR excess.  While a wall
with a higher T$_{wall}$ produces more 10 {\micron} emission, it cannot
account for the excess beyond 20 {\micron}.   As is seen in previous
studies \citep{calvet05, espaillat07a}, a small amount of optically thin
dust can account for the 10 {\micron} emission and the near-IR excess. 
Following \citet{calvet02}, the spectrum for the optically thin region
is calculated as the sum of the emergent flux from optically thin annuli
where the dust in each annulus is heated by stellar radiation.  We hold
the inner radius fixed at the dust destruction radius (0.04 AU) and the
outer radius is determined by the best fit (1 AU). The silicate feature
is well fit with a$_{max}$=4 {\micron} and probes temperatures of $\sim$190--1110 K.  
About 4$\times$10$^{-12}$
${\msun}$ of dust exists in the optically thin region which in the models is composed
of $\sim$80$\%$ amorphous silicates, $\sim$14$\%$ organics, $\sim$3$\%$
amorphous carbon, $\sim$2.6$\%$ troilite,  and less than 1$\%$ enstatite
and forsterite.  The total emission of this optically thin region is
scaled to the vertical optical depth at 10 {\micron}, $\tau_{0}$ $\sim$
0.025.

Since there are no far-infrared or millimeter measurements for CVSO 224,
we cannot constrain the contribution to the SED from the outer disk. 
However, previous papers have shown that the wall and optically thin
dust region dominate the mid-infrared flux \citep{calvet05,
espaillat07a, espaillat07b} and so in this first approximation we
neglect the outer disk.  

\section{Discussion \& Conclusions}

CVSO 224 is $\sim$10 Myr old, making it one of the oldest transitional
disks around a classical T Tauri star.  TW Hya is another
$\sim$10 Myr old transitional disk that has been studied in detail.  In both objects, there is an
inner hole of $\leq$10 AU that contains optically thin dust that is
larger than ISM-sized grains \citep{calvet02}.  The $\sim$2 Myr old CS
Cha also has larger grains in its optically thin region
\citep{espaillat07a}.  In contrast, LkCa 15 and GM Aur in Taurus
($\sim$1 Myr old) have small ISM-sized grains in their optically thin
regions
\citep{espaillat07b, calvet05}. These optically thin disks are
possibly the result of the inward drift of small dust grains from the outer disk
\citep{rice06}. 
Larger dust grains in the optically thin
regions of older transitional disks may imply that there is
some correlation between the process that creates the optically thin
dust and dust evolution over time.
Like GM Aur and LkCa 15, CVSO 224 and TW Hya
have no obvious signs of crystalline silicates in their disks.  
This is hard to explain given that
one would expect 10 Myr of processing to result in crystalline
silicates.  This lack of cystallization is in contrast to 
several young full disks in Taurus which show high degrees
of dust processing (Watson et al. 2008).  



CVSO 224 is an ideal target to test the role of photoevaporation in
transitional disks. Photoevaporation will create an inner hole in disks
with masses $<$0.005 M$_{\sun}$ and mass accretion rates below the
photoevaporative wind rate of $\sim$10$^{-10}$ M$_{\sun}$ yr$^{-1}$
(AA07). CVSO 224's $\mdot$ is below the photoevaporative wind rate,
however, we are detecting a robust mid-infrared excess beyond
20{\micron} from the disk's wall.  This excess agrees with the median
SED of Taurus' T Tauri disks (Figure 1) and is similar to what is observed in
other massive (M$_{disk}$ $>$  0.05 M$_{\sun}$) transitional disks 
(e.g. DM Tau, GM Aur; Calvet et al. 2005), indicating that CVSO 224 still
has a substantial outer disk.  This indication of a massive outer disk
is evidence against photoevaporation being the main cause of
clearing behind the hole we observe in CVSO 224.  Additionally, AA07 show that the size of
the inner hole should reach tens of AU within 0.1 Myr of the onset of
photoevaporation.  Excluding the unlikely possibility that we are
catching CVSO 224 immediately after photoevaporation has switched on,
one would expect the inner hole to be larger and the outer disk to be
mostly gone.  Millimeter observations of CVSO 224 are necessary to
confirm that it has a massive disk in order to eliminate the possibility
that photoevaporation is creating the disk hole around CVSO 224.

If CVSO 224 does have a massive disk
and is not being photoevaporated, one can speculate that the
$\mdot$ of the outer disk, $\mdot_{disk}$, is actually higher than the
observed accretion rate onto the star, $\mdot_{star}$. One way to create
a situation where $\mdot_{star}$ is less than $\mdot_{disk}$ is with
planet formation. If a planet is forming in the disk, the mass coming
from the outer disk is shared by the planet and the inner disk  (Lubow
\& D'Angelo 2006; Najita et al. 2007; AA07). \citet{lubow06} show that
the mass accretion rate past a planet and into the inner disk is 10-25
$\%$ of the mass accretion rate outside the planet's orbit. 
Alternatively, the low mass accretion rate of CVSO 224 could also 
be due to a stellar companion. Hen 3-600
is another weakly accreting $\sim$10 Myr old transitional disk
(7$\times$10$^{-11}$ M$_{\sun}$ yr$^{-1}$; Muzerolle et al. 2000) whose
hole is most likely due to dynamical clearing by its companion
\citep{uchida04}.  Its low $\mdot$ also indicates that photoevaporation
should be clearing this disk, however, Hen 3-600 has a small hole
($\sim$1 AU; Uchida et al. 2004) and its infrared excess \citep{low05}
indicates a substantial outer disk. Hen 3-600's $\mdot_{star}$ could be
limited by binary tidal forces \citep{artymowicz96} and not be
reflective of $\mdot_{disk}$.

CVSO 224 presents an opportune test of photoevaporation and planet
formation in older disks and further observations of this object will
help clarify the role of photoevaporation in disk clearing.  Millimeter
studies of CVSO 224 are necessary to derive the mass of its outer disk. 
If a massive disk is present, its low $\mdot$ could be due to a
planetary or stellar companion and more study will be needed to
distinguish between these two scenarios. 

 \acknowledgments{We thank the referee for comments which helped improve the manuscript.  
 We also thank W. K. M. Rice for discussions.
This work is based on Spitzer Space Telescope data.  N.C. and
and L.H. acknowledge support
from NASA Origins Grants NNG05GI26G and NNG06GJ32G.  N.C. acknowledges JPL grant AR50406.  
P.D. acknowledges
grants from CONACyT, M\`exico. D.M.W. acknowledges support from the
Spitzer Infrared Spectrograph Instrument Project.}

\end{document}